\documentclass{ctr_summer}

\usepackage{ctrfont}
\usepackage{natbib}
\usepackage{undertilde}
\usepackage{color}
\usepackage{cjhebrew}

\usepackage{graphicx}
\usepackage{psfrag}
\usepackage{epsfig}
\usepackage{amsmath,rotating}
\usepackage{dashrule}
\usepackage{undertilde}
\usepackage{comment}

\usepackage{url}

\usepackage{color}
\usepackage{amssymb}

\usepackage{bm}

\title{On interpretability and proper latent decomposition of autoencoders}

\shorttitle{Interpretability and proper latent decomposition}

\author{L. Magri\footnote[1]{Department of Aeronautics, Imperial College London, United Kingdom}\footnote[2]{The Alan Turing Institute, United Kingdom} 
\and N. A. K. Doan\footnote[3]{Faculty of Aerospace Engineering, Delft University of Technology, Netherlands} 
}

\shortauthor{Magri \& Doan}

\begin{document}
\pagenumbering{gobble}

\setcounter{page}{1}

\maketitle

\begin{abstract}
The dynamics of a turbulent flow tend to occupy only a portion of the phase space at a statistically stationary regime. 
From a dynamical systems point of view, this portion is the attractor. 
The knowledge of the turbulent attractor is useful for two purposes, at least:
(i) We can gain physical insight into turbulence (what is the shape and geometry of the attractor?), and 
(ii) it provides the minimal number of degrees of freedom to accurately describe the turbulent dynamics. 
Autoencoders enable the computation of an optimal latent space, which is a low-order representation of the dynamics. If properly trained and correctly designed, autoencoders can learn an approximation of the turbulent attractor, as  shown by~\citet{doanctrsp}. In this paper, we theoretically interpret the transformations of an autoencoder. 
First, we remark that the latent space is a curved manifold with curvilinear coordinates, which can be analyzed with simple tools from Riemann geometry. 
Second, we characterize the geometrical properties of the latent space. 
We mathematically derive the metric tensor, which provides a mathematical description of the manifold. 
Third, we propose a method--- proper latent decomposition (PLD)---that generalizes proper orthogonal decomposition of turbulent flows on the autoencoder latent space. This decomposition finds the dominant directions in the curved latent space. 
This theoretical work opens up computational opportunities for interpreting autoencoders and creating reduced-order models of turbulent flows.

\end{abstract} 


%
%
\section{Introduction}
We write the turbulent problem as a dynamical system, 
and explain the turbulent attractor and its relevance to the latent space of an autoencoder. 
An incompressible turbulent flow is governed by the Navier-Stokes and continuity equations. These are partial differential equations whose solutions, in most realistic regimes, are  spatiotemporally chaotic (turbulent). From a dynamical systems viewpoint, we can      
numerically discretize the spatial variables to define a nonlinear dynamical system
\begin{align}
\frac{d\mathbf{q}}{dt} = \mathsfbi{F}(\mathbf{q}), \quad\quad \mathbf{q}(t=0)=\mathbf{q}_0,
\end{align}
where the nonlinear operator, $\mathsfbi{F}$, encapsulates the boundary conditions.
%
%
%
The Navier-Stokes equations are the mathematical expression of conservation of momentum and mass. 
Conservation laws do not depend on the coordinate system; in other words, they are invariant. 
In more technical terms, we say that they are covariant. 
However, we routinely describe Navier-Stokes in Cartesian coordinates (unless there are clear symmetries, like cylindrical coordinates). 
This is the natural choice because our brain naturally and easily sees width, height, and depth. 
From a geometric point of view, we can think of each grid point as an independent coordinate in the phase space.  
Specifically, we can think of the flow system as a velocity vector in the phase space, where each component is the value of the velocity at a grid point. 
The system's state is represented by the discretized velocity, which is a vector-valued function $\mathbf{q}(t)\in\mathbb{R}^N$, which, geometrically,  is a curve parameterized with time. Physically, $\mathbf{q}(t)$ is a trajectory in the phase space of the flow state (velocity).  
Therefore, $\mathbb{R}^N$ is the phase space of the turbulent flow (in a simple grid and numerical scheme, $N=N_g\times3$, where $N_g$ is the number of grid points and 3 is the number of velocity components). 
Because a turbulent flow is a dissipative dynamical system, the flow state (velocity) tends to occupy a portion of the phase space as the transient dynamics die away (Sturm-Liouville theorem).
This means that the long-term dynamics evolve on an attractor, which can be approximated as an immersed submanifold in $\mathbb{R}^M\subset\mathbb{R}^N$,  where $M<N$. This dimension $M$ can be estimated conservatively with the Lyapunov exponents through the Kaplan-Yorke dimension. 
The turbulent attractor is a curved manifold, for which curvilinear coordinates are best suited to representing the flow state. 
As analyzed in the next section, the autoencoder  can learn a set of curvilinear coordinates that describe the turbulent attractor (assuming that the autoencoder is correctly trained).  

%
%
\section{Autoencoder} 
\label{sec:AE_coord} 

\begin{figure}
    \centering
    \includegraphics[width=0.9\textwidth]{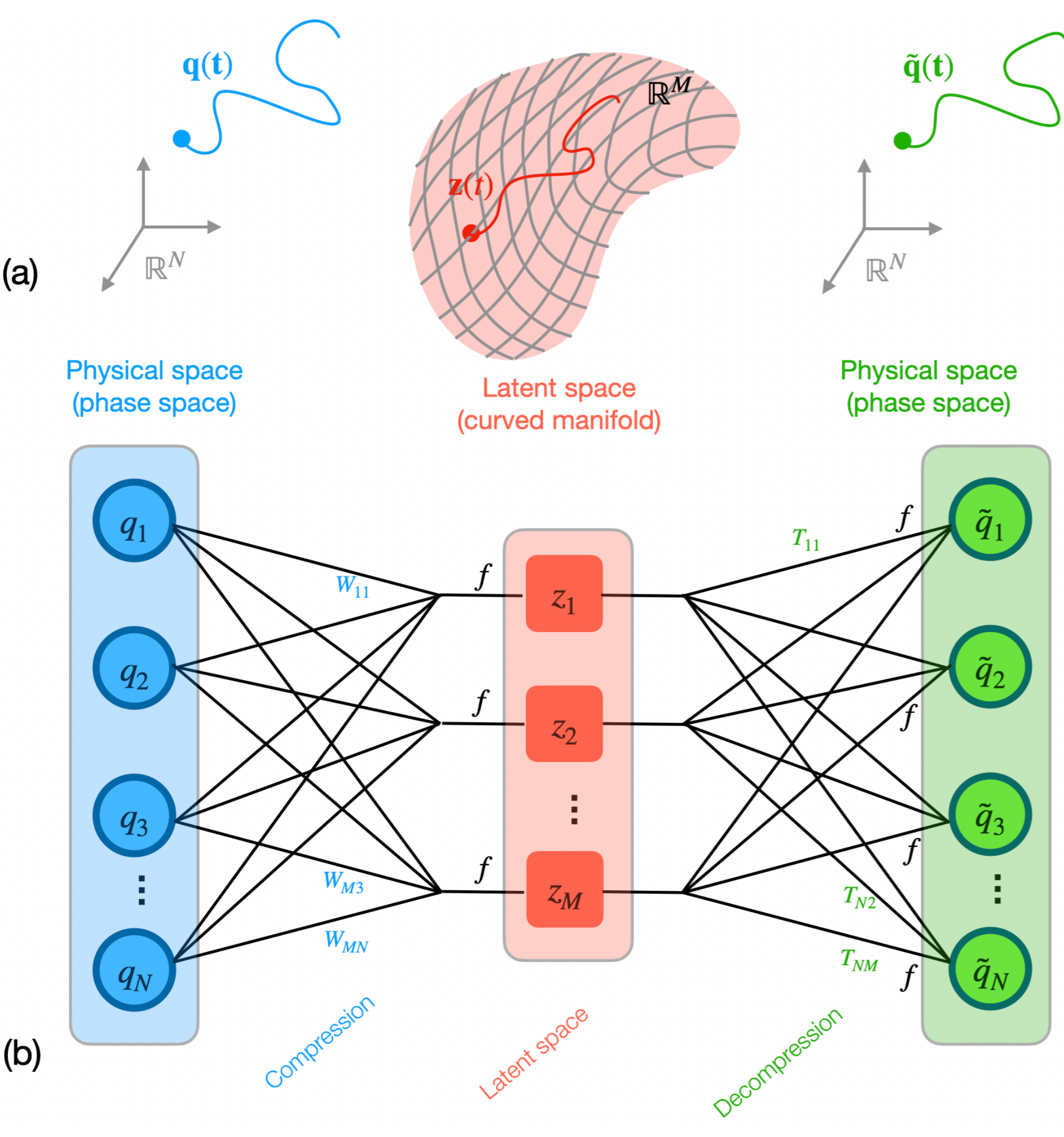}
    \caption{(a) The physical problem that we want to solve. We want to find a low-order  representation (the latent space trajectory, $\mathbf{z}$) of the high-dimensional turbulent flow, $\mathbf{q}$, such that the reconstruction error, $\sim||\mathbf{q}-\tilde{\mathbf{q}}||^2$, is minimized. The latent space is a curved manifold. (b) How the problem is solved with an autoencoder. For brevity, the autoencoder is feedforward, has only one hidden layer (the latent space), and has no bias. }
    \label{Figure_1}
\end{figure}

An autoencoder is an approximation of the identity operator (Figure~\ref{Figure_1}). 
First, it maps the velocity from the physical domain, 
$\mathbf{q}(t)\in\mathbb{R}^N$, 
to a manifold, $\mathbf{z}(t)\in\mathbb{R}^M$, in which, in general, some information is lost.
This is the compression phase (also known as downsampling), which is performed by the encoder.  
The dimension of the latent space, $M$, is user defined. 
Second, it maps the latent space back to the physical space, which provides an approximation of the input vector, i.e., $\tilde{\mathbf{q}}\approx\mathbf{q}$. 
This is the decompression phase (also known as upsampling), which is performed by the decoder.
No information is lost, or gained, in the decompression.   
In the limit of $M=N$, a correctly trained and designed autoencoder learns exactly the identity operator. 
For brevity, we consider a simple autoencoder based on feedforward neural networks with one hidden layer, which is the latent space, and we do not consider biases in the network. These assumptions simplify the mathematical expressions with no loss of generality in the interpretability and conclusions of this paper. 
Mathematically, we consider an autoencoder as 
\begin{align}
\textrm{Encoder:}\quad
\mathbf{z}(t) &= \tilde{f}(\mathsfbi{W}\mathbf{q}(t)), \quad\quad 
\mathbf{z}(t)\in\mathbb{R}^M,  \;\;\; \nonumber\\
z_i &= f\left(\sum_{j=1}^N W_{ij}q_j\right), \label{eq:enco_eq2} \\
\textrm{Decoder:}\quad
\tilde{\mathbf{q}}(t) &= \tilde{f}(\mathsfbi{T}\mathbf{z}(t)), \quad\quad 
\mathbf{\mathbf{q}}(t)\in\mathbb{R}^N,  \;\;\; \nonumber\\
\tilde{q}_i &= f\left(\sum_{j=1}^M T_{ij}z_j\right), \label{eq:deco_eq2} 
\end{align}
where $f(\cdot)$ is the activation function, which is a nonlinear scalar function;  
$\tilde{f}$ is a compact representation of the entry-wise action of $f$; and
the wide matrix, $\mathsfbi{W}$, and tall matrix, $\mathsfbi{T}$, contain the neurons' weights, which are found by solving an optimization problem 
\begin{align}
\mathrm{Find}\;\;\ \mathsfbi{W}\;\textrm{and} \;\mathsfbi{T}\;\;\; \textrm{such that} \nonumber \\ 
\textrm{MSE} &=\frac{1}{N_T}\sum_{n=1}^{N_T} || \mathbf{q}(t_n) - \tilde{\mathbf{q}}(t_n) ||_2^2 \nonumber \\
&=\frac{1}{N_T}\sum_{n=1}^{N_T} || \mathbf{q}(t_n) -  \tilde{f}\left(\mathsfbi{T}\tilde{f}(\mathsfbi{W}\mathbf{q}(t_n)\right) ||_2^2 \nonumber \\ 
&\textrm{is minimum}. \label{eq:optimiz_tra}
\end{align}
Here, MSE stands for mean-squared error, 
 $|| \cdot ||_2$ is the Euclidean norm,   
 and $N_T$ is the number of flow fields that are available in the dataset. 
The optimization problem~\eqref{eq:optimiz_tra} is referred to as training in  neural network jargon. Typically, this is a non-convex optimization problem, which is locally solved by a gradient-based algorithm. 
The components, ${q}_i$, measure the coordinates of the velocity with respect to the chosen reference system for $\mathbb{R}^N$, which is Cartesian in this paper.  
The components $z_i$ measure the coordinates of the velocity vector in the latent space, $\mathbf{z}$, which, as we will see, is a curved manifold. Therefore, $z_i$ are curvilinear coordinates. 

\subsection{Autoencoder as a nonlinear reduced-order model}
In reduced-order modeling, 
we want to find an approximation of the solution, $\tilde{\mathbf{q}}$, such that
\begin{align}
\mathbf{q}(\mathbf{x},t)  = \tilde{\mathbf{q}}(\mathbf{x},t)  + \boldsymbol{\varepsilon}(\mathbf{x},t) , 
\end{align}
where $\boldsymbol{\varepsilon}$ is the approximation error.
A good reduced-order model is such that the error $\boldsymbol{\varepsilon}$ is minimized (with respect to a metric). 
For example, the solution can be approximated by separation of variables as 
\begin{align}
\tilde{\mathbf{q}}(\mathbf{x},t) = \sum_{i=1}^M a_i(t) \boldsymbol{\mathbf{\Phi}}_i(\mathbf{x}), 
\end{align}
where 
$a_i(t)$ are the $M$ velocity components measured with respect to the basis $\boldsymbol{\mathbf{\Phi}}_i(\mathbf{x})\in\mathbb{R}^N$. 
Common choices of the basis vectors, $\boldsymbol{\mathbf{\Phi}}_i(\mathbf{x})$, are the principal components of the data's covariance matrix of the flow field (plus the mean flow). 
From a reduced-order modeling point of view, a neural network autoencoder provides a more general ansatz  
\begin{align}
\tilde{\mathbf{q}}(\mathbf{x},t) 
= 
\sum_{i=1}^M z_i(t) \boldsymbol{\mathbf{\Psi}}_i(z_i(t), \mathbf{x}), \label{eq:dec_non}
\end{align}
where $z_i$ are the  components of $\tilde{\mathbf{q}}$ on the curvilinear basis 
$\boldsymbol{\mathbf{\Psi}}_i(z_i(t), \mathbf{x})$, which is nonlinear and may be non-orthogonal. 
The decomposition in Eq.~\eqref{eq:dec_non} approximates the dynamics nonlinearly; in other words, an infinitesimal variation of the velocity vector results in different variations of the coordinates depending on the position on the submanifold, $d\tilde{\mathbf{q}}=\sum_{i=1}^M dz_i(t) \boldsymbol{\Psi}_i + z_i(t) d\boldsymbol{\Psi}_i$. 
The objective of the autoencoder is to find the optimal basis $\boldsymbol{\Psi}_i(z_i(t),\mathbf{x})$ that minimizes the approximation error (with respect to a metric). 
\section{The geometry of the latent space}
We interpret mathematically the geometry of the latent space with basic tools from differential geometry. 

\subsection{Parametric surface}
The decoder defines a parametric surface, with curvilinear coordinates ${z}_i$. This parametric surface is an $M$-dimensional submanifold embedded in the $N$-dimensional physical space, which is defined as 

\begin{align} \label{eq:parsuf}
\tilde{\mathbf{q}} & = 
\tilde{q}_1(z_1,z_2,\ldots, z_M)\hat{\mathbf{e}}_1 + 
\tilde{q}_2(z_1,z_2,\ldots, z_M)\hat{\mathbf{e}}_2 + 
\ldots 
+
 \tilde{q}_N(z_1,z_2,\ldots, z_M)\hat{\mathbf{e}}_N \nonumber \\ 
& = 
f\left(\sum_{j=1}^M T_{1j}z_j\right)\hat{\mathbf{e}}_1 + 
f\left(\sum_{j=1}^M T_{2j}z_j\right)\hat{\mathbf{e}}_2 + 
\ldots 
+ f\left(\sum_{j=1}^M T_{Nj}z_j\right)\hat{\mathbf{e}}_N \nonumber \\ 
& =  \sum_{i=1}^N f\left(\sum_{j=1}^M T_{ij}z_j\right) \hat{\mathbf{e}}_i, 
\end{align}
This physically defines the submanifold on which the turbulent dynamics converge at a statistically stationary regime (to numerical error). 
Now that we have the parametric equations of the submanifold, we can characterize its geometry, as in the following subsections.

\subsection{Metric tensor}
In the previous sections, we learned that the autoencoder transforms the velocity coordinates from one reference frame, e.g., Cartesian, to a curvilinear frame, i.e., the latent space. 
To analyze the latent space, which is a curved manifold, we can use the decoder to define the parametric equations of the latent space embedded in the physical space (Eq.~\eqref{eq:parsuf}). 
Intuitively, we can think of the decoder as the Rosetta stone to understanding the latent space in a space with which we are familiar, i.e., the physical space (here Cartesian). 
The next question is: How can we investigate the geometry of the latent space? 
We can gain physical insight into the turbulent attractor by characterizing the geometry of the latent space.  
In coordinate transformations, there are quantities that are invariant. 
For example, the squared norm of the differential change in the flow state, $||d\tilde{\mathbf{q}}||^2$, is an invariant:  
It is a number, which does not depend whether we express $\tilde{\mathbf{q}}$ in Cartesian coordinates or in the latent space parameterization. 
The metric tensor is the key quantity to  characterize a curved manifold, which is defined through the invariance of  
\begin{align} \label{eq:metrictensorG}
d\tilde{\mathbf{q}}\cdot d \tilde{\mathbf{q}} 
= 
\sum_{i=1}^N\sum_{j=1}^N d\tilde{q}_i \delta \tilde{q}_j\delta_{ij} 
= \sum_{i=1}^M\sum_{j=1}^M
\underbrace{\frac{\partial \tilde{\mathbf{q}}}{\partial z_i}}_{\equiv\mathbf{g}_i}\cdot \underbrace{\frac{\partial \tilde{\mathbf{q}}}{\partial z_j}}_{\equiv\mathbf{g}_j} dz_i dz_j, 
\end{align} 
where $\delta_{ij}$ is the Kronecker delta, which ensues from the orthonormality of the Cartesian basis vectors.
The vectors $\mathbf{g}_k$ are the basis vectors of the curvilinear system, which vary from one point of the manifold to another. These are a local basis. 
The components of the metric tensor are defined as ${G}_{ij}\equiv \mathbf{g}_i\cdot\mathbf{g}_j$. Intuitively, the metric tensor is the mathematical object that ensures that physical quantities (scalars) such as infinitesimal lengths, angles and areas remain  invariant; that is, their values do not change if the coordinate system changes. 
In the autoencoder, the metric tensor is 

 \begingroup
\allowdisplaybreaks
\begin{align}
%
%
G_{ij} &\equiv \underbrace{\frac{\partial \tilde{\mathbf{q}}}{\partial z_i}}_{\equiv\mathbf{g}_i}\cdot \underbrace{\frac{\partial \tilde{\mathbf{q}}}{\partial z_j}}_{\equiv\mathbf{g}_j}  
 \nonumber\\ &= 
 \frac{\partial}{\partial z_i}\left( \sum_{s=1}^N \hat{\mathbf{e}}_s f\left( \sum_{k=1}^M T_{sk}{z}_k\right)\right)
 \cdot 
  \frac{\partial}{\partial z_j}\left( \sum_{p=1}^N \hat{\mathbf{e}}_p f\left( \sum_{l=1}^M T_{pl}{z}_l\right)\right)
  \nonumber\\ 
     \nonumber\\ &= 
   \left( \sum_{s=1}^N \hat{\mathbf{e}}_s \sum_{k=1}^M \frac{\partial f}{\partial \left(\sum_{k=1}^M T_{sk}{z}_k\right)} \frac{\partial \left(T_{sk}{z}_k\right)}{\partial z_i}\right) 
   \cdot 
      \left( \sum_{p=1}^N \hat{\mathbf{e}}_p \sum_{l=1}^M \frac{\partial f}{\partial \left(\sum_{l=1}^M T_{pl}{z}_l\right)} \frac{\partial \left(T_{pl}{z}_l\right)}{\partial z_j}\right) 
        \nonumber\\ &= 
   \left( \sum_{s=1}^N \hat{\mathbf{e}}_s \sum_{k=1}^M f'_s \frac{\partial \left(T_{sk}{z}_k\right)}{\partial z_i}\right) 
   \cdot 
      \left( \sum_{p=1}^N \hat{\mathbf{e}}_p \sum_{l=1}^M f'_p \frac{\partial \left(T_{pl}{z}_l\right)}{\partial z_j}\right) 
           \nonumber\\ &= 
   \left( \sum_{s=1}^N \hat{\mathbf{e}}_s \sum_{k=1}^M f'_s T_{sk}\frac{\partial {z}_k}{\partial z_i}\right) 
      \cdot 
         \left( \sum_{p=1}^N \hat{\mathbf{e}}_p \sum_{l=1}^M f'_p T_{pl}\frac{\partial {z}_l}{\partial z_j}\right) 
              \nonumber\\ &= 
   \left( \sum_{s=1}^N \hat{\mathbf{e}}_s \sum_{k=1}^M f'_s T_{sk}\delta_{ik}\right) 
   \cdot 
      \left( \sum_{p=1}^N \hat{\mathbf{e}}_p \sum_{l=1}^M f'_p T_{pl}\delta_{jl}\right) 
                 \nonumber\\ &= 
   \underbrace{\left( \sum_{s=1}^N \hat{\mathbf{e}}_s f'_s T_{si}\right)}_{\equiv\mathbf{g}_i}
   \cdot 
         \underbrace{\left( \sum_{p=1}^N \hat{\mathbf{e}}_p f'_p T_{pj}\right)}_{\equiv\mathbf{g}_j} 
                 \nonumber\\ &= 
\sum_{s=1}^N 
\sum_{p=1}^N \hat{\mathbf{e}}_s\cdot\hat{\mathbf{e}}_p f'_s T_{si}  f'_p T_{pj}
                 \nonumber\\ &= 
\sum_{s=1}^N 
\sum_{p=1}^N \delta_{sp} f'_s T_{si}  f'_p T_{pj}
                 \nonumber\\ &= 
\sum_{s=1}^N 
 f'_s T_{si}  f'_s T_{sj}. 
\end{align}
 \endgroup
The metric tensor $G_{ij}$  is positive definite; that is,  $\sum\sum G_{ij} dz_i dz_j>0$ for any non-trivial $dz_i dz_j$. This tensor is also known as the Riemann tensor, and the latent space is, thus, a Riemann manifold. 
Crucially, the metric tensor allows us to compute  Christoffel symbols (the connection), geodesics, and covariant differential operators. Therefore, the metric tensor is a key quantity that describes the turbulent latent space that is learned by the autoencoder. 

\section{Proper latent decomposition}
We have derived the metric tensor, $\mathbf{G}$, of the latent space learned by the autoencoder. 
We propose a decomposition method, which generalizes principal orthogonal decomposition (POD) for flows in the latent space. 
POD has the following steps:
(i) Collect $Q$ flow realizations, $\mathbf{q}^{1,2,...,Q}$; 
(ii) compute the sample mean, 
$\bar{\mathbf{q}} = 1/Q \sum_{i=1}^Q \mathbf{q}^{i}$; 
(iii) compute the sample covariance matrix around the mean, 
$\mathbf{C} = 1/(Q-1) \sum_{i=1}^Q (\bar{\mathbf{q}} - \mathbf{q}^{i})\otimes (\bar{\mathbf{q}} - \mathbf{q}^{i})$, where $\otimes$ is the dyadic product; and 
(iv) eigenvalue decompose the covariance matrix, $\mathbf{C} = \mathbf{U}\boldsymbol{\Lambda}\mathbf{U}^T$. 
The eigenbasis, $\mathbf{U}$, is a set of orthogonal vectors, which are named principal components.
The principal components are ranked with their corresponding eigenvalue, $\lambda_i$.  
The larger the eigenvalue, the more the contribution of the principal component to the flow energy.
POD is widely used because it is relatively simple to compute, and the energy ranking provides an interpretation of the flow structures (proper modes).  
However, as is relevant to the goal of this paper of reconstructing the turbulent attractor, POD has limitations because  
(i) it is an orthogonal global basis, which means that the latent space captured by POD can only be affine (a hyperplane), which means that no curved manifold can be efficiently described by POD; 
(ii) the sample mean of POD is not necessarily a solution of Navier-Stokes; and 
(iii) the proper orthogonal modes are not necessarily solutions of Navier-Stokes~\citep[e.g,][]{jcp}.  
To generalize POD, we exploit the fact that the autoencoder can learn a curved manifold (the latent space). The generalization requires tools from Riemann geometry, which are discussed in the next subsections. 

\subsection{Mean on the latent space}
The computation of the mean in the phase space, which is a Cartesian space, is straightforward.
It is less so in the latent space, which is a curved manifold. 
In reference to Figure~\ref{Figure_1}(a), the mean on the latent space is (an approximation of) the expected value of the flow state. 
To generalize to the autoencoder latent space, we first need to recall what a mean is. 
The mean is the point in an ensemble of points that minimizes  the sum-of-squared distance function 
\begin{align}
\boldsymbol{\mu} = \arg \min_{\mathbf{q}_i\in \mathcal{Z}} d(\boldsymbol{\mu}, \mathbf{q}_i)^2, 
\label{eq:mean_manifold}
\end{align}
where $\mathcal{Z}$ is the latent space, and 
$d$ is the distance function. In Cartesian spaces, $\mathcal{Z}=\mathbb{R}^M$ and $d(\mathbf{x}, \mathbf{y})^2\equiv || \mathbf{x} - \mathbf{y} ||^2$, for any $\mathbf{x}$ and $\mathbf{y}$. 
On the other hand, on manifolds, $d(\mathbf{x}, \mathbf{y})$ is a geodesic. 
The mean in Eq.~\eqref{eq:mean_manifold} is also known as the Fr\'echet mean.
The mean on the manifold in Eq.~\eqref{eq:mean_manifold} requires an optimization problem to be solved. 
This can be achieved by gradient descent, as detailed by~\citet{pennec2019riemannian}. 
Because the latent space can be a good representation of the turbulent attractor (providing that the autoencoder is correctly designed with a lower bound for the number of latent variances equal to the Kaplan-Yorke dimension), the mean on the latent space is a physical realization of the flow state (to numerical error). The sample mean on a manifold is not unique. Think of a sphere: There are infinite means for two points that are diametrically opposite on the sphere. Although not unique, it is unlikely that pathological situations of non-uniqueness appear in high-dimensional turbulent problems. Now that we have the mean on the latent space, we can compute the covariance on the latent space, as detailed in the next section. 

\subsection{Covariance on the latent space}
Similarly to the mean, the sample covariance relies on the definition of a distance function between two points. 
Geometrically, we make an approximation to make the problem more tractable. 
We project the data from the latent space onto the tangent plane centered at the mean. 
In other words, we assume that most of the data are concentrated around the mean;  thus, a tangent approximation can be motivated. 
The sample covariance centered around the mean on the latent space is 
\begin{align}
\boldsymbol{\Xi} = 
\frac{1}{Q-1} 
\sum_{i=1}^{Q} 
\textrm{Log}_{\boldsymbol{\mu}}\mathbf{q}_i  
\otimes 
\textrm{Log}_{\boldsymbol{\mu}}\mathbf{q}_i, 
\end{align}
where $\textrm{Log}$ is the logarithmic map of the manifold~\citep{frankel2011geometry}, which takes a point on the manifold, $\mathbf{q}_i$ and projects it to the tangent plane centered at $\boldsymbol{\mu}$.  

\subsection{Proper latent decomposition}
\begin{figure}
    \centering
    \includegraphics[width=0.9\textwidth]{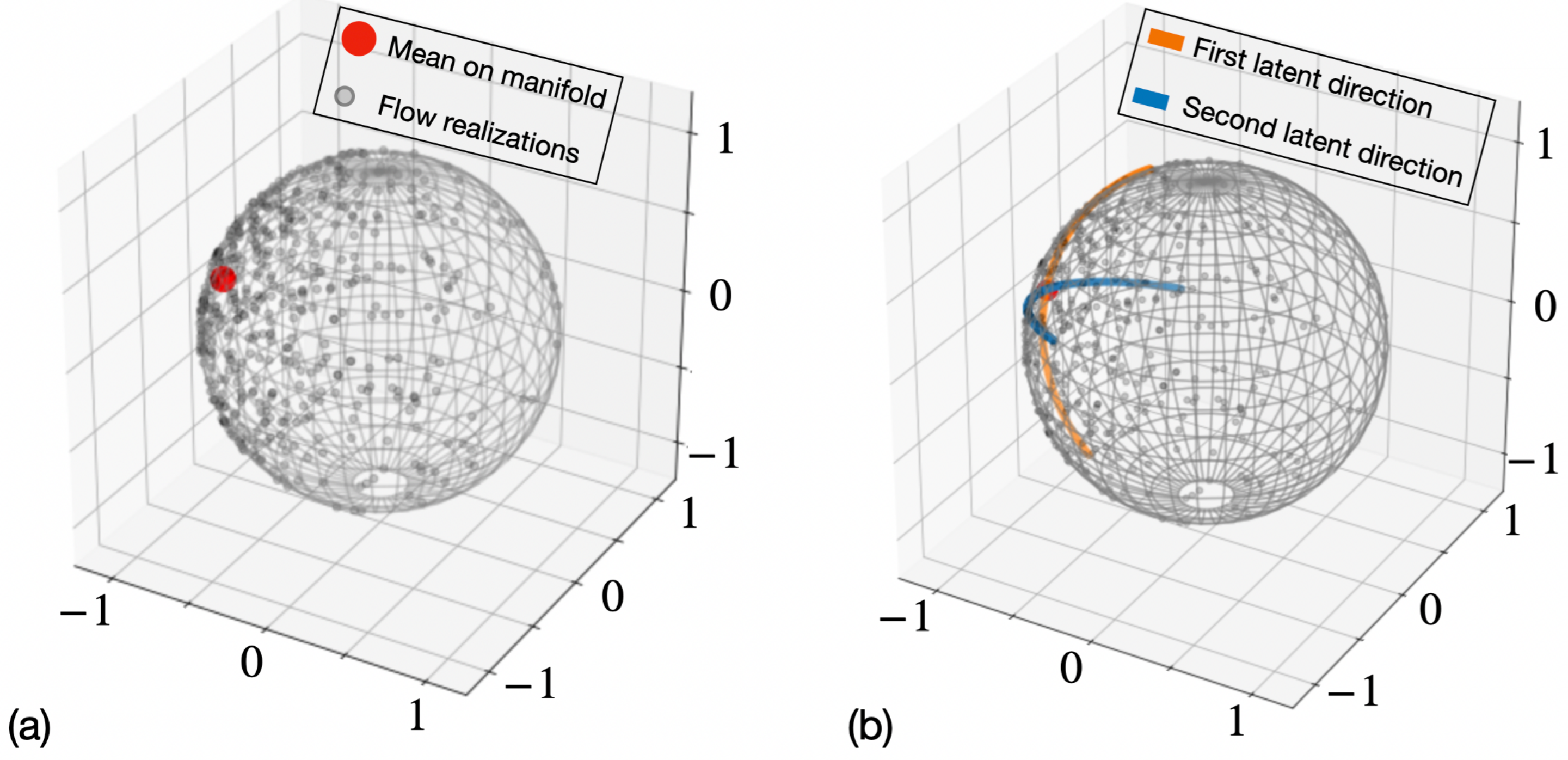}
    \caption{Proper latent decomposition (PLD) on a simple manifold, i.e., the sphere on which the data was randomly generated. (a) Black dots are random data, which represent (pedagogically) the flow realizations.  
    The red dot indicates the mean on the latent space. (b) Principal latent directions computed around the mean.  }
    \label{fig:PLD}
\end{figure}

We can now apply principal component analysis on the sample covariance. 
Because the covariance matrix is defined in an affine space (the tangent plane at the mean), we do not need to modify anything in the eigenvalue decomposition 
\begin{align}
\boldsymbol{\Xi} = \mathbf{U}\boldsymbol{\Lambda}\mathbf{U}^T.
\end{align}
This gives the principal directions on the tangent plane at the mean.
To project these direction back in the latent space, we formally use the exponential map~\citep{frankel2011geometry} 
\begin{align}
\mathcal{U} = \textrm{Exp}_{\boldsymbol{\mu}} \mathbf{U}. 
\end{align}
Therefore, the geodesics, $\mathcal{U}$, are the principal latent directions, which generalize proper orthogonal decomposition. Because they belong in the latent space, which is an approximation of the attractor, they are physical realizations of the Navier-Stokes equations (to numerical error). 
The logarithmic and exponential maps are inverse to each other.
These can be computed from the definition of the geodesics, which can be computed by the covariant derivative, which, in turn, can be computed with the metric tensor and connection. The metric tensor (Eq.~\eqref{eq:metrictensorG}) is the enabler of all the analysis proposed. 
The PLD is shown for a pedagogical problem in Figure~\ref{fig:PLD}. 

\section{Discussion}
We set up a theoretical framework to interpret an autoencoder and to decompose the turbulent flow with the proper latent decomposition (PLD), which generalizes proper orthogonal decomposition to nonlinear manifolds. In current and future work, the research effort will be on translating this theoretical framework into a computational framework, with a thorough assessment of the physical insight gained from the proper latent modes, the computations required for the logarithmic and exponential maps, and the accuracy of the reduced-order models from PLD.

\subsection*{Acknowledgments}
The authors thank Nina Miolane for advice on the computation of geodesics and Riemann geometry.

\bibliographystyle{ctr}

\end{document}